\documentclass[traditabstract]{aa}
\usepackage{natbib,ifthen,setspace,txfonts}
\usepackage{graphicx}
\usepackage{appendix}
\usepackage{lscape}
\begin{document}

\title{The mass-loss return from evolved stars to the Large Magellanic Cloud}
\subtitle{III. Dust properties for carbon-rich asymptotic giant branch stars\thanks{Figures 3 and 4 are only available in electronic form via
http://www.edpsciences.org}}

\date{Received 2010 May 14; accepted 2010 September 12}

\titlerunning{Modeling LMC carbon stars}
\authorrunning{S. Srinivasan et al.}

\author{S. Srinivasan\inst{\ref{inst1}}, B. A. Sargent\inst{\ref{inst2}}, M. Matsuura\inst{\ref{inst3},\ref{inst4}}, M. Meixner\inst{\ref{inst2}}, F. Kemper\inst{\ref{inst5}}, A. G. G. M. Tielens\inst{6}, K. Volk\inst{\ref{inst2}}, A. K. Speck\inst{\ref{inst8}}, Paul M. Woods\inst{\ref{inst5}}, K. Gordon\inst{\ref{inst2}}, M. Marengo\inst{\ref{inst7}}, G. C. Sloan\inst{\ref{inst9}}}   
\institute{Institut d'Astrophysique de Paris, 98 bis Boulevard Arago, Paris 75014, France, email: {\tt srinivas@iap.fr}\label{inst1} \and Space Telescope Science Institute, 3700 San Martin Drive, Baltimore, MD 21218, USA\label{inst2} \and 
UCL-Institute of Origins, Astrophysics Group, Department of Physics and Astronomy, University College London, Gower Street, Kondon WC1E 6BT, United Kingdom\label{inst3} \and UCL-Institute of Origins, Mullard Space Science Laboratory, University College London, Holmbury St. Mary, Dorking, Surrey RH5 6NT, United Kingdom\label{inst4} \and 
Jodrell Bank Centre for Astrophysics, School of Physics \& Astronomy, University of Manchester, Manchester, M13 9PL, UK\label{inst5} \and NASA Ames Research Center, SOFIA Office, MS 211-3, Moffet Field, CA 94035, USA\label{inst6} \and Department of Physics and Astronomy, Iowa State University, Ames, IA 50010, USA\label{inst7} \and Department of Physics and Astronomy, University of Missouri, Columbia, MO 65211, USA\label{inst8} \and Center for Radiophysics and Space Research, Cornell University, 222 Space Sciences Building, Ithaca, NY 14853, USA\label{inst9}}
        
\abstract
{We present a radiative transfer model for the circumstellar dust shell around a Large Magellanic Cloud (LMC) long-period variable (LPV) previously studied as part of the Optical Gravitational Lensing Experiment (OGLE) survey of the LMC. OGLE LMC LPV 28579 (SAGE J051306.40--690946.3) is a carbon-rich asymptotic giant branch (AGB) star for which we have {\it Spitzer} broadband photometry and spectra from the SAGE and SAGE-Spec programs along with broadband {\it UBVIJHK$_{\rm s}$} photometry. By modeling this source, we obtain a baseline set of dust properties to be used in the construction of a grid of models for carbon stars. We reproduce the spectral energy distribution of the source using a mixture of amorphous carbon and silicon carbide with 15\% SiC by mass. The grain sizes are distributed according to the KMH model, with $\gamma=3.5$, $a_{\rm min}=0.01$ $\mu$m\ and $a_0=1.0$ $\mu$m. The best-fit model produces an optical depth of 0.28 for the dust shell at the peak of the SiC feature (11.3 $\mu$m), with an inner radius of about 1430 R$_\odot$ or 4.4 times the stellar radius. The temperature at this inner radius is 1310 K. Assuming an expansion velocity of 10 km~s$^{-1}$, we obtain a dust mass-loss rate of $2.5 \times 10^{-9}$ M$_\odot$ yr$^{-1}$. We calculate a 15\% variation in this mass-loss rate by testing the sensitivity of the fit to variation in the input parameters. We also present a simple model for the molecular gas in the extended atmosphere that could give rise to the 13.7 $\mu$m\ feature seen in the spectrum. We find that a combination of CO and C$_2$H$_2$\ gas at an excitation temperature of about 1000 K and column densities of $3\times 10^{21}$ cm$^{-2}$ and $10^{19}$ cm$^{-2}$ respectively are able to reproduce the observations. Given that the excitation temperature is close to the temperature of the dust at the inner radius, most of the molecular contribution probably arises from this region. The luminosity corresponding to the first epoch of SAGE observations is 6580 L$_\odot$. For an effective temperature of about 3000 K, this implies a stellar mass of 1.5--2 M$_\odot$\ and an age of 1--2.5 Gyr for OGLE LMC LPV 28579. We calculate a gas mass-loss rate of $5.0 \times 10^{-7}$ M$_\odot$ yr$^{-1}$\ assuming a gas:dust ratio of 200. This number is comparable to the gas mass-loss rates estimated from the period, color and 8 $\mu$m\ flux of the source.}


\maketitle

\section{Introduction}
The asymptotic giant branch (AGB) is the final stage in the evolution of low- and intermediate-mass stars (0.8 to 8 M$_\odot$). Products of the nuclear reactions in the AGB star interiors are brought to the outer regions where they form molecules and, farther from the star, dust grains. Carbon-rich (C-rich) AGB stars are created by the process of  the third dredge-up \citep{Iben83}.
Slow winds, with typical expansion velocities of 10 km s$^{-1}$, feed the processed material (gas and dust) back into the surrounding interstellar medium (ISM) at rates up to $10^{-5}-10^{-4}$ M$_\odot$ yr$^{-1}$, where it is eventually locked into newly forming stars. The large numbers of AGB stars makes them significant contributors to the chemical evolution of galaxies. In particular, C-rich AGB stars are the among the major sources of carbon atoms \citep{Dwek1998} and carbonaceous dust grains \citep{Dwek1998,Matsuuraetal2009} in a galaxy. It is therefore crucial to determine the composition of AGB star dust and the total injection rate of this dust into the ISM.  

The study of Galactic carbon stars is complicated due to the high (and often unknown) line-of-sight extinction towards these sources. Despite this fact, a lot of information has been gained about the circumstellar dust around these stars. The Large Magellanic Cloud offers an exceptional environment for the study of the mass-loss return and enrichment process by AGB stars.
Recent near-infrared (NIR) and mid-infrared (MIR) surveys ({\it e.g.}, 2MASS, \citet{Skrutskieetal2006}; DENIS, \citet{Epchteinetal1994} and SAGE, \citet{Meixneretal2006}) have revealed tens of thousands of AGB star candidates in Large Magellanic Cloud (LMC). The SAGE \cite[][]{Meixneretal2006} survey has observed the LMC in the Spitzer Space Telescope \citep[{\it Spitzer};][]{Gehrzetal2007} Infrared Array Camera (IRAC; 3.6, 4.5, 5.8 and 8.0 $\mu$m) and Multiband Imaging Photometer for Spitzer (MIPS; 24, 70 and 160 $\mu$m) bands. \citet{Blumetal2006} and \citet{Srinivasanetal2009} (hereafter, Paper I) used SAGE photometry to identify about 7,\ 000 carbon-rich AGB star candidates in the LMC. Follow-up spectroscopy using the Infrared Spectrograph (IRS) on {\it Spitzer} has also been performed on selected SAGE targets as part of the SAGE-Spec \citep{Kemperetal2010} program. 

Detailed radiative transfer (RT) studies conducted on Galactic and extragalactic carbon stars in the past have helped identify many dust species as possible constituents of their circumstellar envelopes. Here we concentrate on two species that significantly affect the near- and mid-IR photometry and spectroscopy, carbon dust (amorphous carbon or graphite) and silicon carbide ($\alpha$ or $\beta$ type). IRC+10\,216 \citep{NeugebauerLeighton1969} is the best-studied carbon star in the galaxy, and a majority of early studies involved the modeling of its circumstellar shell \citep[{\it e.g.},][]{JonesMerrill1976,Fazioetal1980}. \citet{Rowan-RobinsonHarris1983} used their RT code to fit the circumstellar envelopes (CSEs) of 44 Galactic carbon stars and found that amorphous carbon dust (hereafter AmC) produced better fits than graphite. A similar result was obtained by \citet{MartinRogers1987}. \citet{SkinnerWhitmore1988} found that the strength of the SiC feature in carbon stars was inversely proportional to the total mass-loss rate which they explained as being due to a decrease in the SiC/AmC ratio with increasing mass-loss rate, thus decreasing the prominence of the SiC feature in comparison to the AmC continuum. This decrease in SiC/AmC ratio with increasing mass loss was verified by \citet{ChanKwok1990} and explained as being due to carbon star evolution. Similar trends were also reported by \citet{Lorenz-MartinsLefevre1993,Lorenz-MartinsLefevre1994} and by \citet{Groenewegen1995}. \citet{Groenewegen1995}, \citet{Specketal1997} and \citet{Blancoetal1998} compared the results of models using $\alpha$- and $\beta$-SiC, finding that the dust containing $\alpha$-SiC was able to reproduce the spectra of most of their sources. \citet{Bressanetal1998} included a treatment of AGB dust shells in their stellar population models, and they generated isochrones for comparison with IRAS data for Miras and OH/IR stars. \citet{vanLoonetal1999} performed RT modeling on the {\it ISO} spectroscopy of LMC AGB stars to identify the chemistry of the circumstellar dust and calculated their mass-loss rates and luminosities. \citet{Suh2000} obtained empirical opacity functions for amorphous carbon dust based on their RT modeling of IR spectra as well as laboratory-measured optical data. \citet{Groenewegen2006} presented synthetic photometry for O--rich and C--rich AGB stars from the results of RT models that spanned the relevant range of stellar and dust shell parameters seen in Galactic AGB stars. In particular, they considered two kinds of dust species for carbon star dust shells: AmC and a mixture of AmC and 15\% $\alpha$-SiC by mass. The theoretical work of \citet{Mattssonetal2008} and \citet{Wachteretal2008} showed that the mass-loss rates of carbon stars may not be sensitive to metallicity. Numerous recent studies using {\it Spitzer} spectra of AGB stars in low-metallicity Local Group galaxies \citep[{\it e.g.},][]{Sloanetal2006,Zijlstraetal2006,Lagadecetal2007,Matsuuraetal2007a,Sloanetal2008} seem to support this claim. Similar results were obtained by \citet{Groenewegenetal2007}, who found in their modeling of the IRS spectra of 60 carbon stars in the Magellanic Clouds that the trend of mass-loss rates with period of luminosity was comparable to that of Galactic sources. \citet{Groenewegenetal2009} extended this study to a larger sample, including 110 carbon stars, and compared their mass-loss rates and luminosities with those predicted by evolutionary models as well as dynamical wind models.

Our aim is to develop a grid of RT models for dust shells around carbon stars so that we may fit the SAGE photometry of carbon star candidates and derive their mass-loss rates. These mass-loss rates will enable the calculation of the total carbon-star mass-loss return to the LMC. As a first step, in this paper, we determine a representative set of dust properties for LMC C--rich AGB stars to be used as input to the modeling. We present a {\bf 2D}ust\ radiative transfer model for the circumstellar shell around the variable carbon star OGLE LMC LPV 28579 (SAGE J051306.40-690946.3) with the primary goal of deriving the dust properties with a physically realistic model of the source, as constrained by available data. We have also used this approach in Sargent et al. 2010 (in press) for the O--rich AGB stars and red supergiants. This paper is organized as follows. We describe  the observational data for OGLE LMC LPV 28579 (hereafter LPV 28579) from various studies  in Sect. \ref{sec:cagbmodel:obs}. In Sect. \ref{sec:cagbmodel:analysis}, we provide details of our radiative transfer model for the circumstellar dust, as well as a simple model for the molecular features observed in the spectrum. We discuss the results of the models in Sect. \ref{sec:cagbmodel:discuss}.

\section{Observations}
\label{sec:cagbmodel:obs}

The choice of LPV 28579 for this study was motivated by the availability of both SAGE photometry and SAGE-Spec spectroscopic data. The SAGE-Spec program selected a bright subsample of the SAGE AGB candidates for good quality spectra. This requirement means that the spectroscopic sample is biased towards redder colors, and that LPV 28579 is redder than most carbon stars in the LMC. Nevertheless, it is only moderately optically thick, and exhibits features found in typical carbon stars. Its spectrum (see Sect. \ref{subsec:cagbmodel:obs:sagespec} for details) shows the overall continuum emission from carbon dust as well as a significant 11.3 $\mu$m feature due to SiC, which makes it a good testbed for the dust properties around LMC carbon stars. \citet{Groenewegen2004} classified the star as an obscured AGB candidate, while in Paper I we labelled it an ``extreme" AGB star based on the \citet{Blumetal2006} selection criterion ($J-[3.6]>3.1$ mag). Examination of the SAGE-Spec spectrum (Section \ref{subsec:cagbmodel:obs:sagespec}) shows the 11.3 $\mu$m\ silicon carbide feature to be in emission. Moreover, results from the present modeling work (Section \ref{subsubsec:cagbmodel:parvar}) suggest at best a moderate 11.3 $\mu$m\ optical depth. Based on this information, LPV 28579 would not be considered an extreme carbon star \citep{Specketal2009}.

Photometry for LPV 28579 is available from many recent LMC surveys \citep[{\it e.g.},][]{Zebrunetal2001, Epchteinetal1999, Cutrietal2003, Meixneretal2006, Katoetal2007}, enabling us to constrain its spectral energy distribution (SED).  We have combined the SAGE data  with photometry from the optical Magellanic Clouds Photometric Survey \citep[MCPS;][]{Zaritskyetal1997} as well as the 2 micron All Sky Survey \citep[2MASS,][]{Skrutskieetal2006}. We discuss the optical and NIR variability observations in Sect. \ref{subsec:cagbmodel:obs:variability}, the SAGE photometry in Sect. \ref{subsec:cagbmodel:obs:sagephot} and the SAGE-Spec data in Sect. \ref{subsec:cagbmodel:obs:sagespec}. In Fig. \ref{fig:cagbmodel:variability}, we show the full range of photometry and spectroscopy values measured to illustrate the source variability.  In this paper, we only model the SAGE Epoch 1 photometry, using the optical and NIR data along with the IRS spectrum to constrain the shape of the resulting SED.

\subsection{Variability in Optical and NIR Data}
\label{subsec:cagbmodel:obs:variability}

LPV 28579 was observed as part of the Optical Gravitational Lensing Experiment \citep[OGLE-II;][]{Zebrunetal2001} survey of the Magellanic Clouds. The {\it I}\ band light curve is available as part of the OGLE-II variable star catalog \citep{Szymanski2005} as well as the OGLE-III list of long-period variables (LPVs) as described in \citet{Soszynskietal2009}. Based on its light curve, LPV 28579 was classified as a Mira-type LPV. \citet{Itaetal2004} and \citet{Groenewegen2004} crossmatched the OGLE-II data with the IRSF LMC Survey \citep{Katoetal2007}, DENIS \citep{Epchteinetal1999} and 2MASS All-Sky Release \citep{Cutrietal2003} NIR catalogs and fit the light curves to obtain variability periods and amplitudes. \citet{Itaetal2004} and \citet{Groenewegen2004} calculated the period of variation to be 361 d and 358.6$\pm$0.136 d respectively. \citet{Itaetal2004} also reported a mean {\it I}\ magnitude of 17.483 mag and a peak-to-peak amplitude of 1.865 mag. The OGLE-III catalog lists the mean magnitude as 17.586 mag. The light curve available from the OGLE-III website\footnote{\tt http://ogledb.astrouw.edu.pl/$\sim$ogle/CVS/} shows that LPV 28579 is multi-periodic, with a primary period of 356.2 d (peak-to-peak amplitude 1.587 mag) modulated by a slow variation of 5747 d (peak-to-peak amplitude 0.784 mag). Like many of our extreme AGB candidates identified in Paper I, LPV 28579 is absent from the \citet{Fraseretal2008} MACHO catalog of LMC LPVs. This is most likely due to the moderately high circumstellar extinction at optical wavelengths. The primary period for LPV 28579 is very close to those of the sources in the ``one year artifact" list, but the culling of these artifacts in \citet{Fraseretal2008} was done in a manner as to avoid accidental removal of genuine sources. Based on its brightness ({\it K$_{\rm s}$}=10.9 mag) and period, LPV 28579 would probably fall on period sequence 1, the sequence found by \citet{Woodetal1999} to be predominantly populated by Miras in their fundamental mode of pulsation. While we do not have {\it JHK$_{\rm s}$} light curves, the 2MASS and IRSF photometry for LPV 28579 capture some of the NIR flux variation\footnote{The difference in {\it K}$_{\rm s}$ magnitude between the two epochs is 0.5 mag. Comparing this result to Fig. 3 of \citet{Whitelocketal2003}, which shows the $K$-band pulsation amplitude as a function of period, we find that the amplitude corresponding to the primary period for LPV 28579 is $\sim$0.4--0.8 mag.} (Fig. \ref{fig:cagbmodel:variability}). 

The optical properties of LPV 28579 are fairly normal for LMC carbon stars -- the pulsation period of $\sim$1 yr is average for typical LMC carbon Miras. It has a relatively small amplitude in comparison to other carbon Miras in the OGLE-III catalog (amplitudes ranging from 0.8 mag to 5.7 mag), and this points to a somewhat hotter central star than a very cool one. The moderate amplitude combined with the $\sim$1 yr period make OGLE LMC LPV 28579 a typical member of the LMC carbon Mira family. Furthermore, while LPV 28579 has very red colors ({\it J--K}=3.27 mag), it is nowhere nearly as red as, {\it e.g.}, IRC+10216 \citep[{\it J--K}=6.15 mag;][]{Whitelocketal2006}.

\subsection{SAGE Photometry}
\label{subsec:cagbmodel:obs:sagephot}
\begin{figure}[!tb] 
\resizebox{\hsize}{!}{\includegraphics{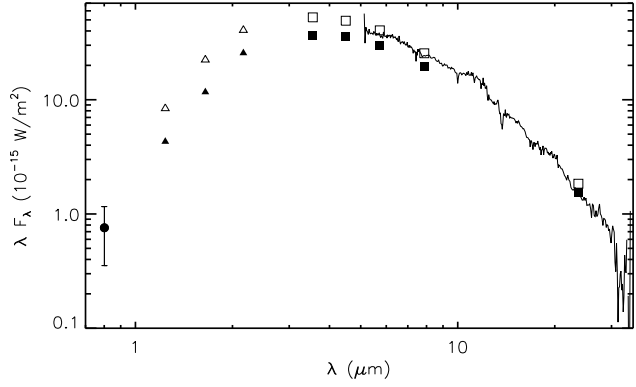}}
\caption{A compilation of photometric and spectroscopic information for LPV 28579. The solid circle and error bar show the mean OGLE {\it I}\ band flux and its range of variation respectively. The 2MASS (open triangles), IRSF (filled triangles) and SAGE (Epoch 1: open squares, epoch 2: filled squares) fluxes are also shown. The solid curve is the SAGE-Spec IRS spectrum.\label{fig:cagbmodel:variability}}
\end{figure}

The SAGE survey \citep{Meixneretal2006} imaged a $\sim 7^\circ \times 7^\circ$ area of the Large Magellanic Cloud (LMC) with the {\it Spitzer} Infrared Array Camera \citep[IRAC;][3.6, 4.5, 5.8 and 8.0 $\mu$m]{Fazioetal2004} and Multiband Imaging Photometer \citep[MIPS;][24, 70 and 160 $\mu$m]{Riekeetal2004}. Two epochs of observations separated by three months were obtained to constrain source variability.  IRAC and MIPS epoch 1 and epoch 2 archive photometry for LPV 28579 is available as part of SAGE data delivered to the Spitzer Science Center\footnote{Data version S13 available on the SSC website, {\tt http://ssc.spitzer.caltech.edu/legacy/sagehistory.html}} (SSC). By comparing data from both epochs of SAGE, \citet{Vijhetal2009} identified infrared variable sources
in the LMC of which $\sim$81\%  are AGB candidates.  LPV 28579  is one of the $\sim$2\,000 SAGE variables discovered by \citet{Vijhetal2009}.   Figure \ref{fig:cagbmodel:variability} shows both epochs of SAGE data.  In this paper, we model the Epoch 1 data (the brighter set of fluxes in Fig. \ref{fig:cagbmodel:variability}) and use the Epoch 2 photometry only to estimate the range of variability of the source (see Sect. \ref{subsubsec:cagbmodel:analysis:centralstar}).

\subsection{SAGE-Spec Data}
\label{subsec:cagbmodel:obs:sagespec}

Spectroscopic follow-up observations to SAGE using the Infrared Spectrograph (IRS) aboard {\it Spitzer} were performed as part of the SAGE-Spectroscopy survey \citep{Kemperetal2010}.  The spectra were reduced using techniques described in the SAGE-Spec Data Delivery Handbook\footnote{Handbook version 1 (30 July 2009) available on the SSC website,\\{\tt http://data.spitzer.caltech.edu/popular/sage-spec/}} \citep{Woodsetal2009} and summarized briefly as follows. LPV 28579 was observed in both spectral orders of the Short-Low (SL; 5.2--14.3 $\mu$m, R$\sim$60--120) and Long-Low (LL; 14.3--13.7 $\mu$m, R$\sim$60--120) instruments on board {\it Spitzer} at two nod positions. Data from the S15.3 and S17.2 pipelines for SL and LL, respectively, were obtained from the SSC for LPV 28579 (AOR \# 22415360). SL sky subtraction was performed by subtracting the observations in one order from the other while keeping the nod position fixed, while LL sky subtraction involved subtracting one nod position from the other while keeping the order fixed. Bad pixels on the detector arrays were replaced with values determined by comparing to the local point-spread function\footnote{For details, please refer to the IRS Data Handbook:\\ {\tt http://ssc.spitzer.caltech.edu/IRS/dh/dh32.pdf}}. 

The width of the apertures used for point source signal extraction increased in proportion to the wavelength; this involved using the profile, ridge and extract modules\footnote{These modules are available in SPICE, the {\it Spitzer} IRS Custom Extraction package.}. Flux densities were obtained by calibrating the raw signal extractions using observations of the standard stars HR 6348 (K0 III), HD 166780 (K4 III) and HD 173511 (K5 III). While all three standard stars were used for LL, only the first was used for SL. Where available, the mean flux from multiple data collection events (DCEs) for a single module/order/nod was computed, with the standard deviation from the mean providing a measure of the corresponding uncertainty.  Spikes not eliminated by replacement of bad pixel values were effectively removed by replacing any uncertainty larger than the average in the local wavelength range by the uncertainty from the nod position closer to the local average. Bonus-order spectra (corresponding to a small segment of the first-order spectrum obtained by the second-order slit) were averaged with spectra from the second and first orders over wavelengths they shared in common.

The IRS spectrum for LPV 28579 (Fig.~\ref{fig:cagbmodel:variability}) shows a broad 11.3 $\mu$m\ SiC emission feature combined with a featureless continuum characteristic of carbonaceous dust. LPV 28579 does not show any particular signs of the 30 micron feature which is strong in the more extreme Galactic carbon stars \citep{HrivnakVolkKwok} and is also seen in very red LMC carbon stars in the SAGE-Spec sample. There is also a 13.7 $\mu$m\ C$_2$H$_2$\ feature. The strength of this feature cannot be reproduced by a model photosphere alone, pointing to its origin in the extended atmosphere and circumstellar shell \citep[see][]{Matsuuraetal2006}. We model both the circumstellar dust as well as the molecular gas in the extended atmosphere in Sect. \ref{sec:cagbmodel:analysis}. We note two trends at longer wavelengths: there is a step-down in the flux at $\sim$20.5 $\mu$m, and the flux errors for the entire Long-Low 1st order (LL1) wavelength range are on average very large. These are suggestive of sky-subtraction issues, so we checked the raw data for LL1. We find that there is a noticeable amount of extended emission in LL1 due to an extremely red source creeping into the slit, which affects the background subtraction.

The optical, NIR and MIR photometry as well as MIR spectroscopy described above are incorporated into Fig. \ref{fig:cagbmodel:variability}. The two-epoch photometry shows that there is no significant change in the shape of the MIR SED. We have taken advantage of this fact and scaled the SAGE-Spec spectrum down to the SAGE Epoch 1 5.8 $\mu$m\ flux to enable easy comparison to photometry. This scaling changes the flux predicted for the spectrum in the 5.8 $\mu$m\  band by $\sim$5\%.

\section{Analysis}
\label{sec:cagbmodel:analysis}
\subsection{2Dust Radiative Transfer Models}
We model the emission from LPV 28579 using the two-dimensional radiative transfer code {\bf 2D}ust\ \citep{UetaMeixner2003}. For simplicity, we assume a spherical geometry for this study. {\bf 2D}ust\ is capable of radiative transfer of axisymmetric systems. This may be needed for describing the most evolved AGB stars as well as their post-AGB counterparts in the model grid. Although AGB mass loss appears to be largely spherical, the successors to these objects, the post-AGB stars, are expected to be bipolar (as are their Galactic counterparts). The dust around post-AGB stars is similar to those around the most evolved AGB stars, therefore we allow for the model grid to cover post-AGB stars and pre-planetary nebulae (PPNe).

The goal of the present study is to extract a set of dust properties for LPV 28579 that will also provide a range of reasonable properties for our entire C--rich AGB sample.  As input, {\bf 2D}ust\ needs information about the central star (effective temperature, size, distance and SED), the shell geometry (inner and outer radius, density variation), and the dust grains (species, optical depth at a reference wavelength, grain size distribution). Table \ref{tab:cagbmodel:twodustparam} summarizes the model parameters that define the star+shell system. Several of these parameters are very sensitive to the modeling process and for these we have shown a best fit parameter, as well as a range of acceptable fits, in the table. Other parameters are adopted based on reasonable assumptions because of limited choices, ancillary data or because the modeling process is not very sensitive to their choice. We describe the selection of these parameters and the estimation of the parameter range in the subsections below.


\subsubsection{Central Star}
\label{subsubsec:cagbmodel:analysis:centralstar}
We model the central star using the carbon star model photospheres of \citet{Gautschy-Loidletal2004}, placing it at a distance of 50 kpc \citep[LMC distance; see, {\it e.g.},][]{Feast1999,vanLeeuwenetal2007}.  We prefer these model spectra over a simple blackbody, as the inclusion of molecular absorption changes the photospheric features in the optical and NIR, which in turn affects the output spectrum (a 15--20\% difference in IRAC fluxes between the 3000 K blackbody and the 3000 K model photosphere discussed below). The \citet{Gautschy-Loidletal2004} set includes 29 solar-mass, solar-metallicity models. As photospheric molecular features are sensitive to metallicity \citep[{\it e.g.},][]{Matsuuraetal2002}, we may not be able to faithfully reproduce their strengths in LMC stars \citep[$Z/Z_\odot\sim$0.3--0.5 for the LMC;][]{Westerlund1997}, but such differences for LPV 28579 would be overwhelmed by emission from its optically thick dust shell. The near-IR photometry suffers from circumstellar extinction, making it difficult to constrain the effective temperature and surface gravity. Therefore, we are only able to find a suitable model by using the source luminosity as a constraint, which we estimate to be (7690$\pm$400) L$_\odot$ for the source using the Epoch 1 fluxes. Two models have luminosities of 6090 L$_\odot$, within 20\% of our estimate for the star. They have effective stellar temperatures of 3000 K and 3100 K, and $\log{g}$ values of $-0.55$ and $-0.49$ respectively. To model LPV 28579, we use the 3000 K photosphere. Keeping all other parameters fixed, we found that there is very little difference in the near- and mid-IR output SEDs if we picked the 3100 K model instead. The model photosphere has a C/O ratio of 1.3. While this value is typical for Galactic Miras, the C/O ratio may be higher for Magellanic Cloud carbon stars \citep[see, {\it e.g.},][]{Matsuuraetal2005,Stanghellinietal2005}, which would change the optical depth due to increased molecular opacity. However, since we are limited by the small number of model photospheres available, we ignore these effects.

\subsubsection{Circumstellar Shell Geometry}
\label{subsubsec:cagbmodel:analysis:shellgeometry} 
We make the simplifying assumption of spherical symmetry as we are more interested in constraining the dust properties in this study. We also assume a constant mass-loss rate and outflow velocity $\upsilon_{\rm exp}$, leading to an inverse-square density distribution in the shell. 
The lower metallicity of the LMC suggests lower $\upsilon_{\rm exp}$ values for the O--rich AGB outflows compared to their Galactic counterparts, but the outflow velocities of carbon stars are less sensitive to metallicity \citep{Habing1996}. \citet{Wachteretal2008} compute carbon-star dynamical wind models for sub-solar metallicities. They find that the outflow velocities of solar metallicity models are about twice that of the LMC. These theoretical predictions of lower outflow velocities at lower metallicity are supported by the observations of \citet{Lagadecetal2010}. If the expansion velocity is assumed to be independent of distance in the shell, it has no effect on the shape of the output SED and cannot, therefore, be constrained by performing radiative transfer. For simplicity, we assume a value of $\upsilon_{\rm exp}$=10 km s$^{-1}$, which is in agreement with observations of Galactic C--rich AGB stars of similar periods \citep[see Fig. 4 of][]{Marigoetal2008}.  The mass-loss rate depends linearly on $\upsilon_{\rm exp}$ (see Eq. \ref{cagbmodel:MLR}) for our simplifying set of assumptions, therefore the calculated rate will be lower if the outflow velocity is lower than the Galactic value.

Radiative transfer codes like DUSTY \citep{Nenkovaetal2000} accept the dust temperature at the inner radius as input and calculate the inner radius on execution, and this temperature is typically fixed at the condensation temperature of the dust species. This condition is not necessarily true for the highly evolved AGB stars as well as post-AGB stars. {\bf 2D}ust adopts a more general approach  \citep[see][]{UetaMeixner2003} by using the inner radius to calculate the temperature structure in the shell. For sources with detached shells, the inner radius can be directly observed in high-resolution images.  The shape of near- and mid-IR SED constrains the inner radius and therefore the temperature at this radius as well. The inner radius of the shell need not, therefore, correspond to the condensation radius, and the resulting temperature may not be equal to the condensation temperature of the dust.

The ratio $R_{\rm max}/R_{\rm in}$ of the outer radius of the dust shell to the inner radius is kept fixed at 1000. The outer radius determines the total shell mass, which in turn is related to the mass-loss timescale. While this timescale is an important quantity, we are more interested in obtaining the  AGB mass-loss rate, which is only weakly sensitive to changes in the outer radius as long as it is large enough to include contributions to the flux at the longest wavelengths of interest (in our case, the 24 $\mu$m\ band) from the coldest dust in the shell. Like \citet{vanLoonetal2003}, we find that the output SED is insensitive to changes in $R_{\rm max}$ out to $\sim$30 $\mu$m. The mass-loss rate is, however, very sensitive to the value chosen for the inner radius. Observations of Galactic carbon stars \citep[{\it e.g.}, IRC+10216,][]{Danchietal1990,Danchietal1995} and results of radiative transfer modeling \citep[{\it e.g.},][]{vanLoonetal1999, Groenewegenetal1998} suggest inner radii in the range $\sim$2--20 stellar radii ($R_*$), while simple estimates based on energy balance suggest that amorphous carbon dust should form within a few $R_*$ \citep{Hofner2007}. We vary the inner radius within these bounds until the desired SED shape is obtained.

\subsubsection{Dust Grain Properties}
\label{subsubsec:analysis:2dust:dprop}
We model the dust around LPV 28579 using amorphous carbon \citep[optical constants from the ``ACAR" sample of][]{Zubkoetal1996} and silicon carbide \citep[optical constants from][]{Pegourie1988} grains in order to reproduce the strong 11.3 $\mu$m\ SiC feature and long-wavelength continuum observed.

Our grain sizes are distributed according to the \citet{Kimetal1994} (KMH) prescription\footnote{$n(a)\propto a^{-\gamma}\exp{[-a/a_0]}$ with $a>a_{\rm min}$}, with a power-law falloff of index $\gamma$=3.5. We vary the minimum grain size $a_{\rm min}$ and the scale height $a_0$ until the shape of the spectrum is reproduced. We find noticeable changes in the output SED only when $a_{\rm min}$ is increased beyond $\sim$0.1 $\mu$m. This insensitivity to the small grain size limit is due in part to the fact that larger grains contribute more significantly to the dust mass, making it easier to place constraints on the scale height than on the minimum grain size \citep[a similar result was obtained by][]{Specketal2009}. Given this distribution of sizes as input, {\bf 2D}ust\ then internally calculates absorption and scattering cross-sections and asymmetry factors for an ``average" spherical dust grain from the Mie theory \citep{Mie1908} assuming isotropic scattering. We run the code in the Harrington averaging \citep{Harringtonetal1988} mode, which gives an average grain size of $\sim$0.1 $\mu$m, the typical value used in single-size models \citep[{\it e.g.},][]{vanLoonetal1999, Groenewegen2006}.

{\bf 2D}ust requires as input the optical depth at a reference wavelength, which we choose to lie near the center of the SiC emission (11.3 $\mu$m). We iteratively adjust this optical depth and relative abundance of SiC until the shape of the feature is satisfactorily reproduced.

\subsubsection{Fitting procedure}
\label{subsubsec:cagbmodel:fitprocedure}
{\bf 2D}ust uses the dust opacities, dust grain and shell properties and the density variation as input to first calculate the density $\rho_{\rm in}$ at the inner radius from the relation
\begin{equation}
\label{cagbmodel:tau}
\tau_\lambda=\kappa_\lambda\int_{R_{\rm in}}^{R_{\rm out}}\rho(r) dr
\end{equation}
where $\tau_\lambda$ is the optical depth at the reference wavelength $\lambda$, and $\kappa_\lambda$ is the opacity averaged over dust size and composition. The total dust mass in the shell is calculated once $\rho_{\rm in}$ is known. This is divided by the wind crossing time in the shell to obtain
\begin{equation}
\label{cagbmodel:MLR}
{\dot M}=4\pi\frac{\tau_\lambda}{\kappa_\lambda}R_{\rm in}\upsilon_{\rm exp}
\end{equation}
The dust mass-loss rate can be calculated once a value for the expansion velocity is chosen.

The overall fitting procedure is summarized as follows. Starting with the model photosphere that comes closest in flux to the luminosity of the source (see Sect. \ref{subsubsec:cagbmodel:analysis:centralstar}) we modify the inner radius, optical depth, SiC abundance and the range of grain sizes until the observed SED as well as the SiC feature strength is reproduced. In order to compare the results of {\bf 2D}ust with our observations, we generate synthetic photometry in the {\it UBVIJHK$_{\rm s}$} as well as IRAC and MIPS24 bands in a manner similar to \citet{Srinivasanetal2009}. At each step, we scale the {\bf 2D}ust output SED to the Epoch 1 8 $\mu$m flux for comparison. The scaling affects the luminosity as well as the mass-loss rate, and in Sect. \ref{subsubsec:cagbmodel:parvar} we investigate the quality of the fit by varying this scale, which results in a range of predicted luminosities and mass-loss rates.

Starting with the $T_{\rm eff}$=3000 K photosphere, we first increase the optical depth until we roughly match the optical extinction as well as the mid-IR continuum observed in the photometry. The inner radius is now varied to improve the fit in the near-IR and IRAC part of the SED (the outer radius is kept fixed at 1000 $R_{\rm in}$, see Sect. \ref{subsubsec:cagbmodel:analysis:shellgeometry}). At this point, we compare the SiC features in the model and the IRS spectrum, and change the relative abundance of SiC in the model if required. In general, this change affects the IRAC fluxes, and we compensate for this by modifying the optical depth and/or inner radius as needed. Finally, we vary the minimum and maximum grain sizes to improve the fit. As noted in Sect. \ref{subsubsec:analysis:2dust:dprop}, changes in the minimum grain size do not significantly affect the spectrum. We terminate the iterative procedure described above when we are able to satisfactorily match the optical, near-IR and IRAC/MIPS24 photometry for LPV 28579, as well as reproduce the 11.3 $\mu$m feature in its IRS spectrum. The resulting best-fit model is described below.

\subsubsection{2Dust Fit Results and Sensitivity To Parameter Variation}
\label{subsubsec:cagbmodel:parvar}
Table \ref{tab:cagbmodel:twodustparam} summarizes the fixed parameters as well as the values of the free parameters corresponding to the best-fit model. The {\bf 2D}ust output spectrum is superimposed on the photometry and IRS spectrum in Fig. \ref{fig:cagbmodel:totalfit}. The luminosity of the model photosphere is lesser than the value calculated for LPV 28579 from photometry, so the entire SED has to be scaled up. The value of this scale factor decides the predicted luminosity and mass-loss rate, as well as the fit quality. We find that we are able to fit the photometry as well as the IRS spectrum for a scale factor of 1.08, which corresponds to a luminosity of about 6580 L$_\odot$. Taking into account the photometric errors, a 1\% to 15\% raise in model luminosity (corresponding to luminosities in the range 6150--7010 L$_\odot$) result in good fits ($\la$ 7\% variation around the best-fit value). To demonstrate that this uncertainty is small compared to the difference in luminosities between the two epochs, we scale the best-fit model down to fit the Epoch 2 photometry and find a luminosity of about 4810 L$_\odot$ ($\sim$27\% fainter). The scaled best-fit dust mass-loss rate for Epoch 1 data is found to be about $2.5\times 10^{-9}$ M$_\odot$ yr$^{-1}$. The optical depth at 11.3 $\mu$m is 0.28. The inner radius of the dust shell is 4.4 $R_*$ (1430 R$_\odot$) and the temperature at this distance from the star is 1310~K.
 
The results of the model fitting are most sensitive to changes in $R_{\rm in}$, the optical depth, the SiC dust abundance and the value of $a_0$. When we have created a grid of models,  we will be able to derive chi-squared errors for the parameters.  For the moment, we determine the acceptable range of variation in the parameters mentioned above as follows. We create boundary ranges on the data by allowing the photometry to change by up to 3 times the uncertainty and by restricting the spectroscopy to a range bounded by its 1$\sigma$ uncertainty, because of lower signal-to-noise ratio associated with the spectra. Using this range of data variation, we determine by eye the range of variation in the model fit due to changes in a certain parameter while keeping all other parameter values fixed at their best-fit model values. The range of parameters thus calculated are shown in Table \ref{tab:cagbmodel:twodustparam}. Fig. \ref{fig:cagbmodel:parvartau} illustrates this process by showing the range in acceptable model fits as a result of changing the optical depth.

The process described above helps us place a rough constraint on the variation of the important output parameters -- the luminosity, mass-loss rate and the dust temperature on the inner radius. We find that changing the inner radius of the dust shell causes the largest variation in both the temperature at the inner radius as well as the mass-loss rate. We find $T_{\rm in}$ values between 1260 K and 1380 K. The mass-loss rate is in the range $(2.4 - 2.9)\times 10^{-9}$ M$_\odot$ yr$^{-1}$. The variation of parameter values thus gives us an estimate of the required data quality if the derived mass-loss rates are to be within $\sim$15\%, assuming our value for $\upsilon_{\rm exp}$ is correct.

\subsection{Modeling The Circumstellar Gas}
\label{sec:cagbmodel:analysis:gas}

The moderate optical depth of the circumstellar shell should be effective in attenuating photospheric features. However, LPV 28579 has strong absorption features around 13.7 $\mu$m. The features therefore originate in the circumstellar region \citep[see, {\it e.g.},][]{Zijlstraetal2006,Matsuuraetal2006}. In Galactic carbon stars, there is also a contribution to this wavelength range due to HCN, but due to the low abundance of nitrogen, this effect is not pronounced in LMC stars \citep{Matsuuraetal2006}.

\begin{figure}[!tb] 
\resizebox{\hsize}{!}{\includegraphics{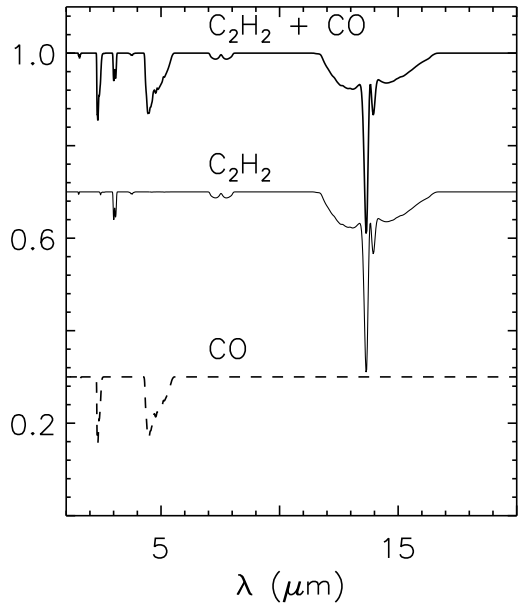}}
\caption{The synthetic spectra for CO (dashed line, column density = $3\times 10^{21}$ cm$^{-2}$) and C$_2$H$_2$\ (thin solid line, column density = $10^{19}$ cm$^{-2}$) from the HITRAN database and \citet{Rothmanetal2009}\label{fig:cagbmodel:gastrace}. The combined spectrum (thick solid line) used to model the molecular contribution to the observed spectrum is also shown. The spectra for CO and C$_2$H$_2$\ are scaled down for clarity.}
\end{figure}

\begin{figure}[!htb] 
\resizebox{\hsize}{!}{\includegraphics{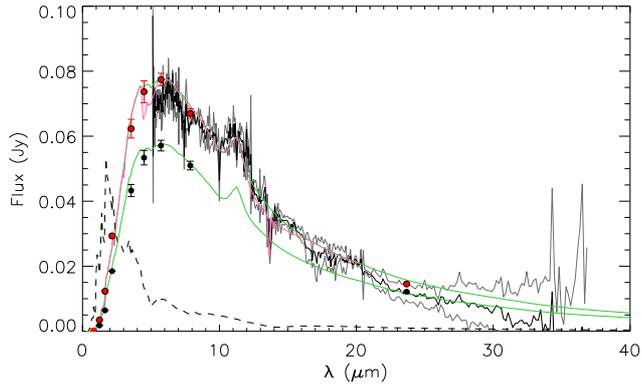}}
\caption{The results of modeling the dust and gas around LPV 28579. The SAGE Epoch 1 (red dots) and epoch 2 (black dots) photometry as well as SAGE-Spec spectrum (black curve) shown with the best-fit 2Dust RT model (green curves) scaled to fit Epoch 1 and Epoch 2 photometry . The photosphere model corresponding to the best-fit SED is also shown (dashed curve). The model SED is convolved with the synthetic molecular spectrum shown in Fig. \ref{fig:cagbmodel:gastrace} to reproduce the 13.7 $\mu$m\ feature (pink curve).\label{fig:cagbmodel:totalfit}}
\end{figure}

\begin{figure}[!htb] 
\resizebox{\hsize}{!}{\includegraphics{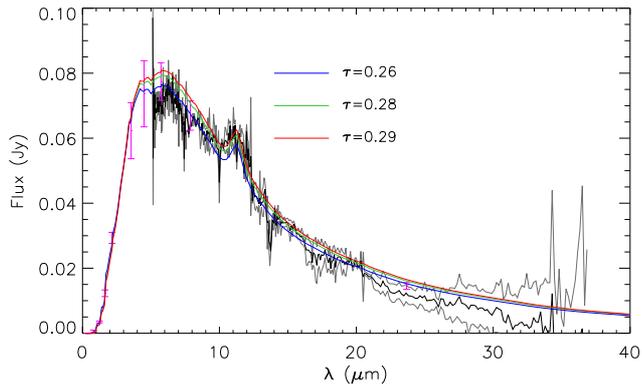}}
\caption{The sensitivity of the {\bf 2D}ust\ model output to optical depth. The SAGE Epoch 1 photometry for LPV 28579 is allowed a 3$\sigma$ variation (magenta error bars), and the spectroscopy is kept to within 1$\sigma$ (gray curves). Model SEDs for 11.3 $\mu$m\ optical depth values spanning this range of allowed fluxes are also shown. The optical depth for good fits ranges from 0.26 (blue curve) to 0.29 (red curve). The best-fit curve for $\tau$=0.28 (green) is also shown for comparison.\label{fig:cagbmodel:parvartau}}
\end{figure}

We simulate the molecular bands of CO and C$_2$H$_2$\ in LPV 28579 using the slab model code of \citet{Matsuuraetal2002} with recent line lists published by \citet{Rothmanetal2009} as part of the HITRAN database. Assuming that the majority of the molecular absorption comes from the circumstellar shell, the excitation temperatures are chosen to be cooler than the temperature at the inner shell radius, {\it i.e.}, $T_{\rm ex}<T_{\rm in}$. Based on these considerations, we derive excitation temperatures of 1000 K for both species, with column densities of $10^{19}$ cm$^{-2}$ and $3\times 10^{21}$ cm$^{-2}$ for C$_2$H$_2$\ and CO respectively. Fig. \ref{fig:cagbmodel:gastrace} shows the locations of the spectral features for the parameters above. The resulting (gas+dust) fit is in good agreement with the spectrum and is plotted in Fig. \ref{fig:cagbmodel:totalfit}. The CO band is at the edge of the IRS spectrum so we do not have observational constraints on it, but we are able to reproduce the 13.7 $\mu$m feature using C$_2$H$_2$. We note here that the model is useful only for identifying the molecular bands and to roughly estimate the excitation temperatures and densities of these molecular species, it can not be used to determine the details of the atmospheric structure. A determination of the mass-loss rate would also require knowledge of the location of the slab of C$_2$H$_2$ in the circumstellar envelope. As this quantity is very uncertain, we do not calculate a gas mass-loss rate using this method.

\section{Discussion}
\label{sec:cagbmodel:discuss}

The long-term goal of this study is to use {\bf 2D}ust\ modeling of carbon star circumstellar shells to fit the photometry of the $\sim$7\,000 C--rich SAGE AGB candidates, along with any carbon stars found among the $\sim$1\,400 sources in the SAGE extreme AGB list, in order to derive their mass-loss rates and to estimate the total mass-loss return from AGB stars to the LMC ISM. We are also deriving the dust properties for O--rich AGB stars in a parallel study (Sargent et al. 2010, in press). In this section, we weight the simplifying assumptions used in our modeling of LPV 28579 against this future goal. Our study of the variation in model parameters demonstrates that we place reasonable constraints on the inner shell radius, the temperature at this radius, dust grain properties and the resulting mass-loss rates. We discuss the results of our modeling in detail below.

\subsection{Dust Properties}

Figure \ref{fig:cagbmodel:totalfit} illustrates the good agreement between our model using an AmC+SiC dust composition and the overall continuum as well as the 11.3 $\mu$m feature observed in the spectrum. Our combined gas+dust model improves on the {\bf 2D}ust\ prediction by fitting the 13.7 $\mu$m absorption. We note two minor discrepancies: the silicon carbide feature seems broader than the model prediction, and the model fit systematically overestimates the flux at wavelengths longward of $\sim$20 $\mu$m. A broadening of the SiC feature may indicate self-absorption \citep[see, {\it e.g.},][]{Specketal1997} corresponding to a high optical depth. The long-wavelength disagreement is probably due to the poor background subtraction (see Sect. \ref{subsec:cagbmodel:obs:sagespec}) rather than due to a modeling issue. Given the low signal-to-noise of the spectrum, it is not certain if these discrepancies are real; we are therefore unable to justify a detailed refinement of our RT model.

Our best-fit model predicts a temperature of 1310 K at the inner radius of the dust shell, which is located at 4.4 $R_*$. In Sect. \ref{subsubsec:cagbmodel:parvar}, we estimate the inner edge to lie within (4.0--4.8) $R_*$ (corresponding to temperatures of 1260--1380 K), which is consistent with the theoretical predictions of \citet{Hofner2007} for the condensation of amorphous carbon dust ($T_{cond}$=1500 K, $R_{\rm in}$=3 $R_*$). The range of temperatures are slightly warmer than would be expected for LMC carbon stars of comparable mass-loss rates \citep[see, {\it e.g.}, Fig. 4 and Equation 1 of][]{Groenewegenetal2009}. This may be due to the fact that we are using a warmer central star. Here we are limited by the availability of the colder photospheres of comparable luminosity. Moreover, as already discussed, the moderate optical depth prevents us from using the observed near-IR colors to estimate the central star's effective temperature.

The KMH grain size distribution for the best-fit model is defined by $\gamma$=3.5, $a_{\rm min}$=0.01 $\mu$m\ and $a_0$=1.0 $\mu$m. As noted in Sect. \ref{subsubsec:analysis:2dust:dprop}, the shape of the output SED is sensitive to the size of the largest grains, but does not vary appreciably with changes in the minimum grain size up $\sim$0.1 $\mu$m. About 15--18\% of the mass is contained in grains with diameters exceeding 1 $\mu$m, depending on the minimum grain size. For comparison, only 0.1\% of the dust mass ejected from IRC+10216 ($L\sim$$1.5\times 10^4$ L$_\odot$, $\upsilon_{\rm exp}$=15 km s$^{-1}$, dust MLR $\sim$$8\times 10^{-8}$ M$_\odot$ yr$^{-1}$) is in micron-sized particles \citep{Jura1994}.

The SiC content derived from the best-fit model is comparable to values derived for dusty Galactic carbon stars \citep{Groenewegenetal1998}. The lower metallicity of the LMC would imply a lower silicon abundance in general, and \citet{Groenewegenetal2007} find lower SiC/AmC ratios for the LMC compared to Galactic stars. By studying the effect of varying the SiC abundance on the output SED, we note that the SiC content can be in the range 10--16\%. In this paper, we assume a gas:dust ratio $\Psi$=200. This value was determined for carbon stars in the solar neighbourhood \citep{Jura1986}. An accurate determination of the gas:dust ratio is still to be made for LMC carbon stars. Some observational and theoretical studies \citep[{\it e.g.},][]{Matsuuraetal2005,Wachteretal2008} indicate that the carbon star gas:dust ratio may not be metallicity dependent. In fact, recent studies \citep{Groenewegenetal2007,Wachteretal2008,Mattssonetal2008} have found that the mass-loss rates of carbon stars may not be metallicity dependent. The gas:dust ratio can be estimated from the mass fraction of SiC in the dust if we also know the fraction of the total Si mass that is in the dust. Assuming that the Si abundance scales with metallicity \citep[The LMC silicon abundance is not well constrained; see the discussion in][]{Matsuuraetal2005} we obtain $\Psi$ in the range 200--570. However, we have assumed here that all of the Si goes into the dust. In reality, the fraction of Si in dust depends indirectly on the amount of carbon in the dust. For example, if half of the carbon remaining after CO formation goes into dust, then 8--13\% of the Si is in dust\footnote{Here, we have assumed that carbon and silicon are the only constituents of the dust. A more realistic estimate would include metallic iron dust, which somewhat counteracts the low Si fraction and lowers the gas:dust ratio.}. The situation improves for higher C/O ratio -- for C/O=1.8, the Si dust fraction is 20--34\%. An alternative way to estimate the gas:dust ratio is to use the \citet{vanLoon2000} relationship $\Psi\sim$ Z$^{-1\pm 0.3}$ for carbon stars, which corresponds to $\Psi$=300--1000 for the LMC. Due to the lack of constraints on the various quantities that determine the gas:dust ratio, we use $\Psi$=200 when quoting a total mass-loss rate for LPV 28579, noting that the mass-loss rate could be up to about 5 times larger based on the calculations above.


\subsection{Mass-loss Rate}
Our {\bf 2D}ust model predicts a dust mass-loss rate of  $2.5\times 10^{-9}$ M$_\odot$ yr$^{-1}$. For a gas:dust ratio of 200, the total mass-loss rate is then $5\times 10^{-7}$ M$_\odot$ yr$^{-1}$. This rate is on the low end of the \citet{vanLoonetal1999} mass-loss rates for the brightest, most obscured LMC carbon stars as well as the rates calculated for the \citet{Groenewegenetal2009} sample. The mass-loss rates calculated from the [3.6]--[8.0] and {\it K}--[8.0] colors of LPV 28579 using equations 1 and 2 of \citet{Matsuuraetal2009} are slightly higher ($1.4\times 10^{-6}$ M$_\odot$ yr$^{-1}$ and $2.5\times 10^{-6}$ M$_\odot$ yr$^{-1}$ respectively). However, our dust MLR is consistent with the value determined from the 8 $\mu$m flux -- using the excess--MLR relation for extreme AGB stars from Paper I, we get a rate of $3.9\times 10^{-9}$ M$_\odot$ yr$^{-1}$. Our total rate agrees with the values calculated using the period--MLR relations of \citet{VW93} and \citet{Groenewegenetal1998} ($10^{-7}$ M$_\odot$ yr$^{-1}$ and $7.4\times 10^{-7}$ M$_\odot$ yr$^{-1}$ respectively, for a period of 356.2d).  The value obtained for the mass-loss rate is sensitive to the $\upsilon_{\rm exp}$ adopted (see Eq. \ref{cagbmodel:MLR}). The outflow velocities for Galactic carbon stars are typically 10 km s$^{-1}$, but can be up to about twice as high. The \citet{Wachteretal2008} models predict LMC outflow velocities to be about half that of comparable Galactic sources. This range of velocities introduces an uncertainty of a factor of two in the calculated mass-loss rate.


\subsection{Evolutionary State}
\begin{figure}[!htb] 
\resizebox{\hsize}{!}{\includegraphics{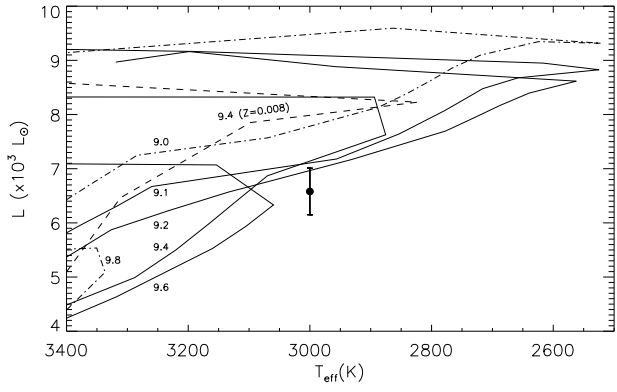}}
\caption{LPV 28579 (black dot with error bar representing range of calculated Epoch 1 luminosities) on the $L-T_{\rm eff}$ diagram. The solid and dot-dashed lines, tagged according to their age ($\log{t/{\rm Gyr}}$), are the \citet{Marigoetal2008} isochrones for $Z$=0.006. For comparison, the $Z$=0.008 track for $\log{t/{\rm Gyr}}$=9.4 is also shown (dashed line).
\label{fig:cagbmodel:evolstate}}
\end{figure}

In order to understand our $L$ and $T_{\rm eff}$ values for LPV 28579 in the context of its evolutionary state, we identify its location in the Hertzsprung-Russel (HR) diagram in relation to the \citet{Marigoetal2008} theoretical isochrones. We choose a metallicity of $Z$=0.006, which is consistent with observational estimates for the metallicity of the LMC. We generate these isochrones using their CMD interactive web interface\footnote{{\tt http://stev.oapd.inaf.it/cmd}} so that we can also extract the initial stellar masses for the stellar tracks. Figure \ref{fig:cagbmodel:evolstate} shows the location of LPV 28579 on the HR diagram. Also shown are selected isochrones for $Z$=0.006 (ages ranging from $\log{t/{\rm Gyr}}$=9.0 to 9.8). We find that the $\log{t/{\rm Gyr}}$=9.1 to 9.6 isochrones are able to produce the range of luminosities we have estimated for LPV 28579 in this study, which gives us an age in the range 1.6--4 Gyr. As noted by \citet{Marigoetal2008}, this is the typical age of the O--rich AGB population that undergoes a smooth transition to C--rich chemistry due to third dredge-up (also see the discussion of the multiple O--rich populations in Paper I). The initial masses corresponding to these isochrones were in the range 1.3--2 M$_\odot$. Figure 3 of \citet{GroenewegendeJong1994} shows the evolution of 1.25 and 5 M$_\odot$\ AGB stars in period-luminosity space. Based on its luminosity and primary period, LPV 28579 falls slightly upward of the 1.25 M$_\odot$\ track, providing a mass comparable with our estimate using theoretical isochrones. 

The derived age (and hence progenitor mass) range is dependent on the metallicity of the isochrones as well as our $T_{\rm eff}$ estimate for LPV 28579. Figure \ref{fig:cagbmodel:evolstate} demonstrates the effect of changing the metallicity (compare the $\log{t/{\rm Gyr}}$=0.94 isochrones for $Z$=0.006 and 0.008). At lower metallicities, the isochrones predict carbon stars at lower luminosities. The optically thick shell of LPV 28579 makes the precise determination of $T_{\rm eff}$ difficult (see \ref{subsubsec:cagbmodel:analysis:centralstar}) so that our choice of model photosphere is made on the basis of the luminosity estimated from photometry. The range of ages and masses from isochrone comparison is dependent on how well we can constrain the effective temperature. If we assume that $T_{\rm eff}$ is cooler than 3100 K, we obtain the age and mass range above. For $T_{\rm eff}$ under 3050 K, the range reduces to $\log{t/{\rm Gyr}}$=9.2--9.5 giving masses in the range 1.4--1.7 M$_\odot$.

The total mass-loss rate of $5.0\times 10^{-7}$ M$_\odot$ yr$^{-1}$\ is comparable to the nuclear-burning rate at the luminosity of the star \citep[see Fig. 9 in][]{vanLoonetal1999}, which suggests that LPV 28579 may undergo the superwind phase at the end of its life on the AGB.

\section{Conclusions}
\label{summary}
We have modeled the circumstellar shell of the carbon-rich asymptotic giant branch star LPV 28579 using the radiative transfer code {\bf 2D}ust\ to fit the SAGE photometry and SAGE-Spec spectroscopy available for the source. Using a mixture of amorphous carbon and silicon carbide dust with 12\% SiC by mass, and dust grains ranging in size from 0.01 $\mu$m\ to $\sim$1.0 $\mu$m, we find a luminosity of 6580 L$_\odot$ and a dust mass-loss rate of $2.5 \times 10^{-9}$ M$_\odot$ yr$^{-1}$. The inner radius of the shell is at 4.4 times the stellar radius and at a temperature of 1310 K. The optical depth at 11.3 $\mu$m\ is found to be 0.28. We model the molecular features in the circumstellar shell with slab models of C$_2$H$_2$\ and CO, and calculate column densities of $10^{19}$ cm$^{-2}$ and $3\times 10^{21}$ cm$^{2}$ for the two species respectively. The model SED when combined with the gas model shows excellent agreement with the observed spectrum and photometry. The total mass-loss rate assuming a gas:dust ratio of 200 is comparable to estimates derived from empirical relations involving the period, color and mid-infrared flux. LPV 28579 has properties typical of dust-enshrouded LMC carbon Miras, and the model derived in this work defines a baseline set of carbon star dust properties in our upcoming model grid.
 \noindent

\begin{acknowledgements}
The authors would like to thank the anonymous referee for their helpful comments on the manuscript. This work is based on observations made with the Spitzer Space Telescope, which is operated by the Jet Propulsion Laboratory, California Institute of Technology under NASA contract 1407. The research in this paper has been funded by Spitzer grant 1310534 and NASA NAG5-12595. The authors have made use of the SIMBAD astronomical database and would like to thank those responsible for its upkeep. The authors {\bf also} thank Bernie Shiao at STScI for his hard work on the SAGE database and his kind assistance.
\end{acknowledgements}

\clearpage
\begin{table*}
\caption{{\bf 2D}ust\ Parameters For LPV 28579}
\label{tab:cagbmodel:twodustparam}
\centering
\begin{tabular}{ll}
\hline\hline\noalign{\smallskip}
\multicolumn{2}{l}{\bf Photosphere model\tablefootmark{a}}\\
\noalign{\smallskip}
T$_{\rm eff}$(K) & 3000\\
$L_*$(L$_\odot$) & 6580 (6150--7010)\tablefootmark{b} (Epoch 1)\\
& 4810 (Epoch 2)\\
$R_*$(R$_\odot$) & 325 (Epoch 1)\\
C/O & 1.3\\[0.25em]
\noalign{\smallskip}
\multicolumn{2}{l}{\bf Dust shell properties}\\
\noalign{\smallskip}
$R_{\rm in}$($R_*$) & 4.4  (4.0--4.8)\\
$R_{\rm{out}}$($R_{\rm in}$) & 1000\\
Density profile & $\rho(r)\sim r^{-2}$\\
$v_{\rm{exp}}$(km~s$^{-1}$) & 10\\
Gas:dust ratio & 200\\[0.25em]
\noalign{\smallskip}
\multicolumn{2}{l}{\bf Dust grain properties}\\
\noalign{\smallskip}
Species & AmC\tablefootmark{c}$+$SiC\tablefootmark{d}\\
SiC fraction & 12\% (10\%--16\%)\\
$\tau$(11 $\mu$m) & 0.27  (0.25 -- 0.275) \\[0.25em]
Size distribution & KMH\tablefootmark{e}\\
 & $a_{\rm min}$($\mu$m) = 0.01 ($<$0.1)\\
 & $a_{0}$($\mu$m) = 1 (0.75--1.3)\\
 & $\gamma$ = 3.5\\
\noalign{\smallskip}
\multicolumn{2}{l}{\bf Mass-loss rate and dust temperature}\\
\noalign{\smallskip}
$\dot{M}_d$($\times$10$^{-9}$ M$_\odot$ yr$^{-1}$) & 2.5  (2.4--2.9) \\[0.25em]
$\dot{M}_g$($\times$10$^{-7}$ M$_\odot$ yr$^{-1}$) & 5.0  (4.8--5.8)\\[0.25em]
$T_{\rm in}$(K) & 1310  (1260 -- 1380)\\
\hline\hline
\end{tabular}
\tablefoot{
\tablefoottext{a}{Photosphere model from \citet{Gautschy-Loidletal2004}.}
\tablefoottext{b}{Uncertainty in parameter value (see Sect. \ref{subsubsec:cagbmodel:parvar} for details).}
\tablefoottext{c}{Amorphous carbon grains, $\rho=1.8$ g cm$^{-3}$, optical constants from \citet{Zubkoetal1996}.}
\tablefoottext{d}{SiC grains, $\rho=3.22$ g cm$^{-3}$, optical constants from \citet{Pegourie1988}.}
\tablefoottext{e}{Size distribution from \citet{Kimetal1994}:\\ $n(a)\sim a^{-\gamma}\exp{\left(-a/a_0\right)}$ with $a>a_{\rm min}$.}
}
\end{table*}

\clearpage

\clearpage

\clearpage

\clearpage

\clearpage
\end{document}